# Dynamics of Gender Bias in Computing

Thomas J. Misa

University of Minnesota

ABSTRACT: Gender bias in computing is a hard problem that has resisted decades of research. One obstacle has been the absence of systematic data that might indicate when gender bias emerged in computing and how it has changed. This article presents a new dataset (N=50,000) focusing on formative years of computing as a profession (1950-1980) when U.S. government workforce statistics are thin or non-existent. This longitudinal dataset, based on archival records from six computer user groups (SHARE, USE, and others) and ACM conference attendees and membership rosters, revises commonly held conjectures that gender bias in computing emerged during professionalization of computer science in the 1960s or 1970s and that there was a 'linear' one-time onset of gender bias to the present (with the erroneous implication that professional computing was inherently, persistently, and permanently gender-biased). Such a linear view also lent support to the "pipeline" model of computing's "losing" women at successive career stages. Instead, this dataset reveals three distinct periods of gender bias in computing and so invites temporally distinct explanations for these changing dynamics. It significantly revises both scholarly assessment and popular understanding about gender bias in computing. It also draws attention to diversity within computing. ACM members likely hold valuable membership and conference records that could clarify the fine structure of women's expanding participation in computing during 1965-85 and the subsequent contraction setting in during the 1990s through today. One consequence of this research for CS reform efforts today is data-driven recognition that legacies of gender bias beginning in the mid-1980s (not in earlier decades) is the problem. A second consequence is correcting the public image of computer science, since this data demonstrates that gender bias is a *contingent* aspect of professional computing, not an intrinsic or permanent one.

In May 1948 women were strikingly prominent in ACM. Founded just months earlier as the "Eastern Association for Computing Machinery," the new professional society boldly aimed to "advance the science, development, construction, and application of the new machinery for computing, reasoning, and other handling of information."[36] No fewer than 27 women were ACM members, and many were leaders in the emerging field.[i] Among them were the pioneer programmers Jean Bartik, Ruth Lichterman, and Frances Snyder of ENIAC fame; the incomparable Grace Murray Hopper who soon energized programming languages; Florence Koons from the National Bureau of Standards and US Census Bureau; and noted mathematician-programmer Ida Rhodes.[26] During the war Gertrude Blanch had organized a massive human computing effort (a mode of computation made visible in the film *Hidden Figures* [2016][47]) and, for her later service to the US Air Force, became "one of the most well-known computer scientists and certainly the most visible woman in the field."[24,25] Mina Rees, a mathematics Ph.D. like Hopper and Blanch, notably funded mathematics and computing through the Office of Naval Research (1946-53), later serving as the first female president of the American Association for the Advancement of Science. In 1949 Rees was among the 33 women (including at least 7 ACM women) who participated in an international conference at Harvard University, chairing a heavyweight session on "Recent Developments in Computing Machinery."[29]

---

[i] The May 1948 ACM membership roster is in Margaret R. Fox Papers (Charles Babbage Institute 45 purl.umn.edu/41420) box 2, folder 9; other ACM rosters in Frances E. Holberton Papers (CBI 94 purl.umn.edu/40810) box 23.

Their prominence has led to the widespread but inaccurate impression that women were numerically dominant in early computer programming. As one account puts it, "at its origins, computer programming was a largely feminized occupation."[18,19] This view, resting on suggestive but fragmentary data, has become prominent in popular culture, scholarship, and mass media, including the *Wall Street Journal* and National Public Radio and the widely acclaimed documentary "Code: Debugging the Gender Gap" (2015) by Robin Hauser Reynolds.[14,41] The film popularized the conjecture by some scholars that "women made up 30% to 50% of all programmers" in the 1950s or 1960s and that male programmers subsequently pushed them out. Porter supports this quote "according to [historian] Ensmenger" (specifically citing the Robin Hauser film).[46]

A recent article[45] in *Communications of the ACM* approvingly cites one such source positing a binary switch from a female-dominated field to a male-dominated one. There Mundy clearly states the linear view: "after World War II, software programming was considered rote and unglamorous, somewhat secretarial—and therefore suitable for women. The glittering future, it was thought, lay in hardware. But once software revealed its potential—and profitability—the guys flooded in and coding became a male realm."[43] It seems widely accepted that men actively remade computer programming from a female- into a male-dominated field during the 1960s or 1970s just as computer science was professionalizing itself through expansion of research, professional societies, and higher education. This accepted view posits that computing was *born female* and then *made masculine*, with a simple linear dynamic leading straight to today's male-dominated profession. One implication of this conjecture is that gender bias was an inherent part of (male-driven) professionalization in computing. In varied forms, "many computer programmers embraced masculinity as a powerful resource for establishing their professional identity and authority," in Ensmenger's formulation.[19]

**The 'linear model' is too simple**

In the absence of systematic data on gender in the computing workforce, prior to the 1970 US Census, such a linear model once seemed plausible.[38] It was furthermore supported by fragmentary and sometimes cherry-picked evidence and buttressed by theoretical claims about the nature of professionalization.[ii] But it is too simple. To start, we need systematic, longitudinal data. For deeper insight on women in computing during these years, this article presents a new dataset with more than 50,000 individuals tabulated by their first (given) names, an indicator of ascribed gender (if not gender identity). The results may be surprising. In 1948 the 27 named ACM women, alongside 330 named ACM men, constituted 7.6% of its membership. Similarly, women were 8.6% and 7.6% of ACM members in 1949 and 1952; and women constituted 7.6% and 5.3% of ACM conference attendees in 1950 and 1952. Women were 5.7% of the 1949 Harvard conference. A retrospective celebration[50] suggests women were 12.7% of the Univac pioneers from 1951 (see Figure 1).[iii] This data does not support the common conjecture that women numerically dominated early computing.

**Figure 1. Women's participation in conferences/ACM members (1948–1953)**

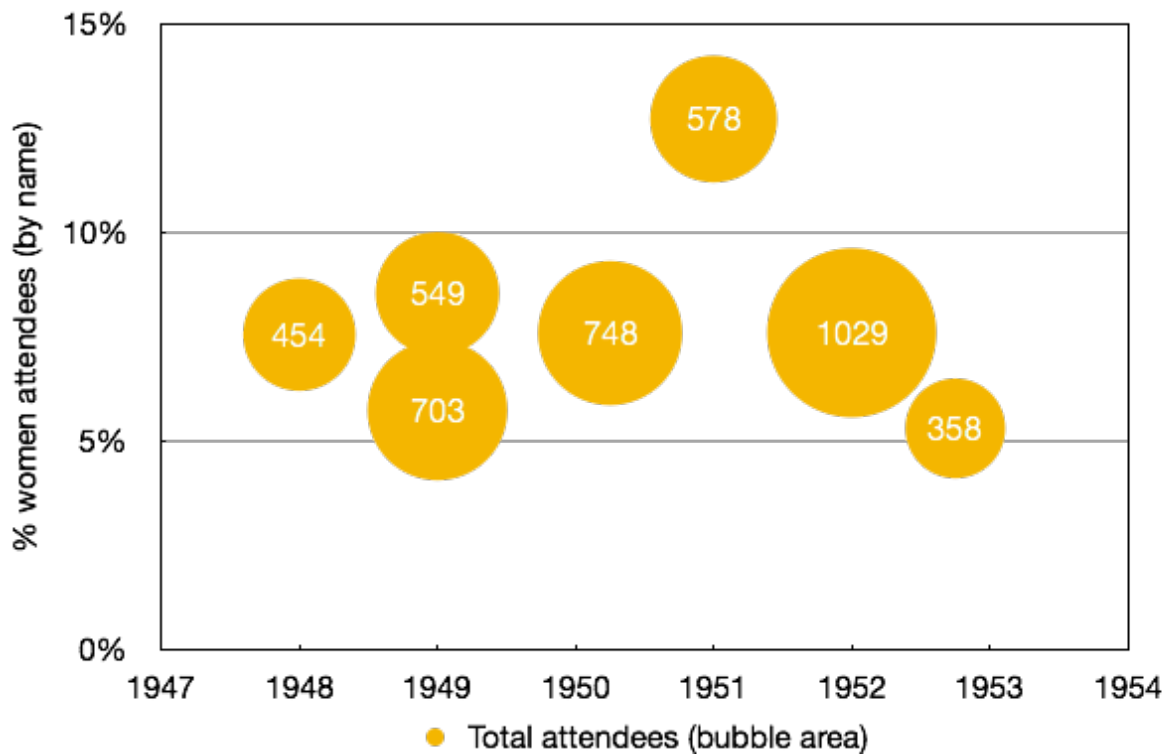

---

[ii] Ensmenger's 1974 source for "reliable contemporary observers"[18,19] claiming 30-plus percent women programmers in fact mentions women on just two pages: a certain single IBM programming group; and a conjecture on women in the "moderating role of 'mother'."[13]

[iii] See UNIVAC Conference 1990, CBI OH 200 at purl.umn.edu/104288; and "NCC 1981 Pioneer Day" at web.archive.org/web/20210108001336/http://lawrencegoetz.com/adr/Univac/doc/Univac_Staff.doc

The "pipeline" model is a related linear view, now widely criticized. In Berryman's influential 1983 Rockefeller Foundation report[5] the pipeline metaphor helped identify the *different* reasons for underrepresentation in the quantitative sciences of African Americans, Hispanics, and American Indians, with structural "losses from the educational pipeline" beginning in high school as well as personal "field choices" (e.g. college major) shaping patterns of underrepresentation. For computer science Camp expanded on Berryman's findings for women that losses were concentrated in a latter stage (from bachelor's to doctoral degrees).[12] In computing the pipeline model posited a one-way decline of women, from the 1980s, noting that the proportion of women "fell" at each career "stage" from undergraduate student through graduate school and on to full professor. Moshe Vardi recently voiced concern about "puncturing the recruiting pipeline."[51]

A recent critique asks: "What's wrong with the pipeline? Everything. The pipeline assumes a passive flow of women (and men) from one stage to the next culminating in a scientific career. Women's underrepresentation in science results then from their leakage from the pipeline."[9] Such a linear model inadequately acknowledges women's diverse career paths and non-academic career stages, better conceptualized as non-linear "pathways." Fox and Kline caution that "women may linger as tenured associate professors without attaining full rank" and so not fully participate in academic decision-making and professional leadership, even while nominally still within the pipeline; in their view the "pathways" model is a better guide to the "dynamic . . . features and forces" of institutional settings, procedures, policies, and cultures in which women faculty members do not always experience orderly, expected, sequential or uni-directional progression through career ranks.[20] Clearly, much more needs to happen than merely "keeping women in the pipeline."[9,52]

To evaluate the 'making programming masculine' thesis and scrutinize the linear–pipeline view, the Charles Babbage Institute analyzed membership and attendee lists of six computer-user groups with available archival records.[41,53] Two of the largest user groups were formed in 1955. SHARE (for IBM computers) and USE (Sperry-Rand Univacs) provided a means for diverse companies, financial institutions, federal agencies and laboratories, and international entities to share algorithms and program code, to identify and address practical problems, to develop novel technical and organizational solutions—and, not least, to give sharp feedback to manufacturers. Both groups compiled attendee lists for their twice-yearly meetings, and many of these list *first names*.

First names, suitably analyzed and methodically tallied, indicate gender; in addition, committee reports identify hundreds of attendees as "Mr" or "Mrs" or "Miss"; oral histories identify others; and the Social Security Administration tabulates all given US birth names by ascribed gender since 1880.[32] Between 80 and 100% of user-group attendees can be gender-identified.[41] Available records also give insight into Control Data's Coop, Burroughs' CUBE,

Digital's DECUS, and the best-selling Mark IV software package for IBM computers. For each user-group, a time-series shows the participation of women in professional computing and indicates the rate of growth. The user-group attendees are taken to be samples of the computing workforce. No single user-group, with the possible exception of SHARE, is anything like a representative sample.

All such historical statistics, including government-compiled ones, are formed from sources of data that vary in uniformity (e.g. direct personal surveys, company personnel reports, trade literature assessments, and industry or trade-group statistics); "uniform data" for historical statistics are always created by researchers, compilers, and analysts.[2,3] This present longitudinal dataset is the largest available for assessing changes in women's participation in the computing workforce (trade journals occasionally conducted one-time salary surveys[27])—until data from the US Census and Bureau of Labor Statistics in the 1970s. The research method introduced here might be used to create longitudinal data, now lacking or fragmentary, on women in the STEM workforce. This systematic approach convincingly supplants earlier studies' reliance on fragmentary data or anecdotal evidence drawn from scattered or non-representative observations.

Figures 2a–f present new time-series data on women's participation in the US computing workforce from 1955 to 1989. Each graph's x-axis gives the years from available archival records;[iv] the y-axis, the percentage of women identified by first names; and the bubble area, the total analyzed population for each year. Individuals with gender-ambiguous or initials-only names are included in the bubble area (N) but are set aside for tabulation of women's participation. The data establishes varied growth across the 1960s and into the 1980s. Women's participation in SHARE grew slowly but steadily from 1955 to 1973, when, with thousands of attendees, it shifted to initials-only names. Women's participation as SHARE officer-managers similarly grew from 1968 to 1989, with a higher $R^2$ value supporting the upward linear-trend line. ($R^2$ is a standard linear regression measure of the 'goodness of fit' of computed trendlines with the underlying data: technically, $R^2$ is the percent of variation in dependent variable [%-women] that can be attributed to variation in independent variable [years]. All trendlines and statistics computed by Mac Numbers 4.3.1.) Notably, after women officer-managers reached 26.5% in 1989, a wider measure of women as SHARE meeting speakers was lower at 16.8% (N=491) and 19.4% (N=443) during 1991-92. Women's participation grew steadily in USE during 1955–89, in CDC's Coop during 1959–64, and in Burroughs' CUBE during 1962–76, all with moderate $R^2$ values. Data from the Mark IV software user group 1969–81, shows strong growth ($R^2$ =0.94) with women's participation reaching 30%.

---

[iv] Archival collections include the Hagley Museum and Library's USE/UNITE records Accession 1881 at findingaids.hagley.org/repositories/3/resources/915 (accessed January 2021) as well as the Charles Babbage Institute's SHARE, USE, Control Data, Burroughs, DECUS, Margaret Fox, and Evan Linick (Mark IV) records at web.archive.org/web/20201002222305/www.cbi.umn.edu/collections/archmss.html

**Figure 2. Women's participation in user groups (1955–1989)**

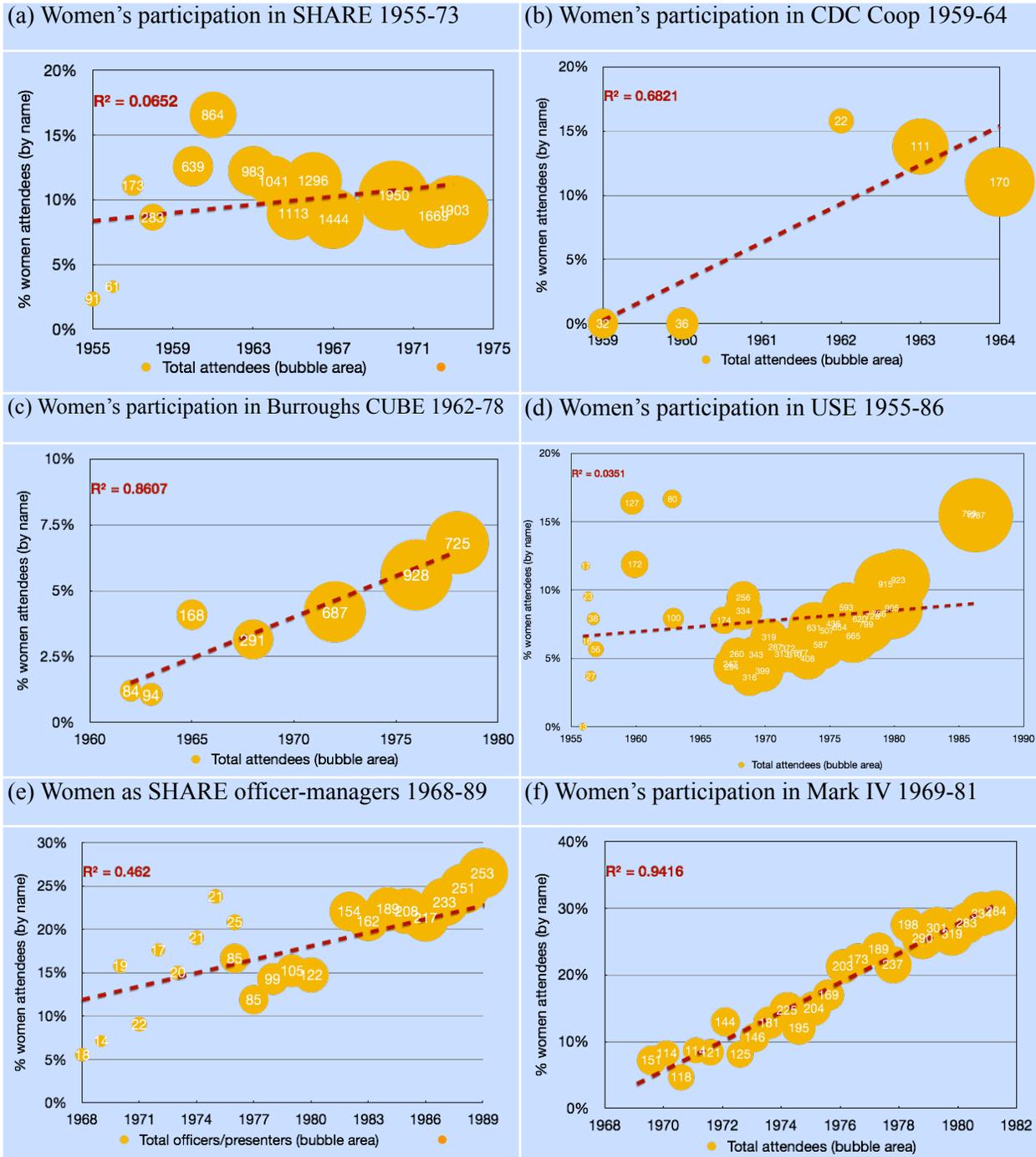

**Gender bias is non-linear**

Figure 3 combines the membership and user-group data across 1948–1995, adding the available federal workforce statistics and US computer-science bachelor degrees from the Bureau of Labor Statistics and NSF, respectively. For clarity, this graph simplifies the time-series data through plotting the underlying trend lines. Figure 3 shows decidedly non-linear dynamics, with varied growth rates and significant declines. The trendlines indicate unmistakable growth 1960s-1980s in women's participation in the computing workforce, refuting the commonly-held "linear model" and any supposed masculine take-over. This user-group data tallies with salary surveys,[v] company-wide group photographs,[vi] and the NSF and BLS/US Census data.

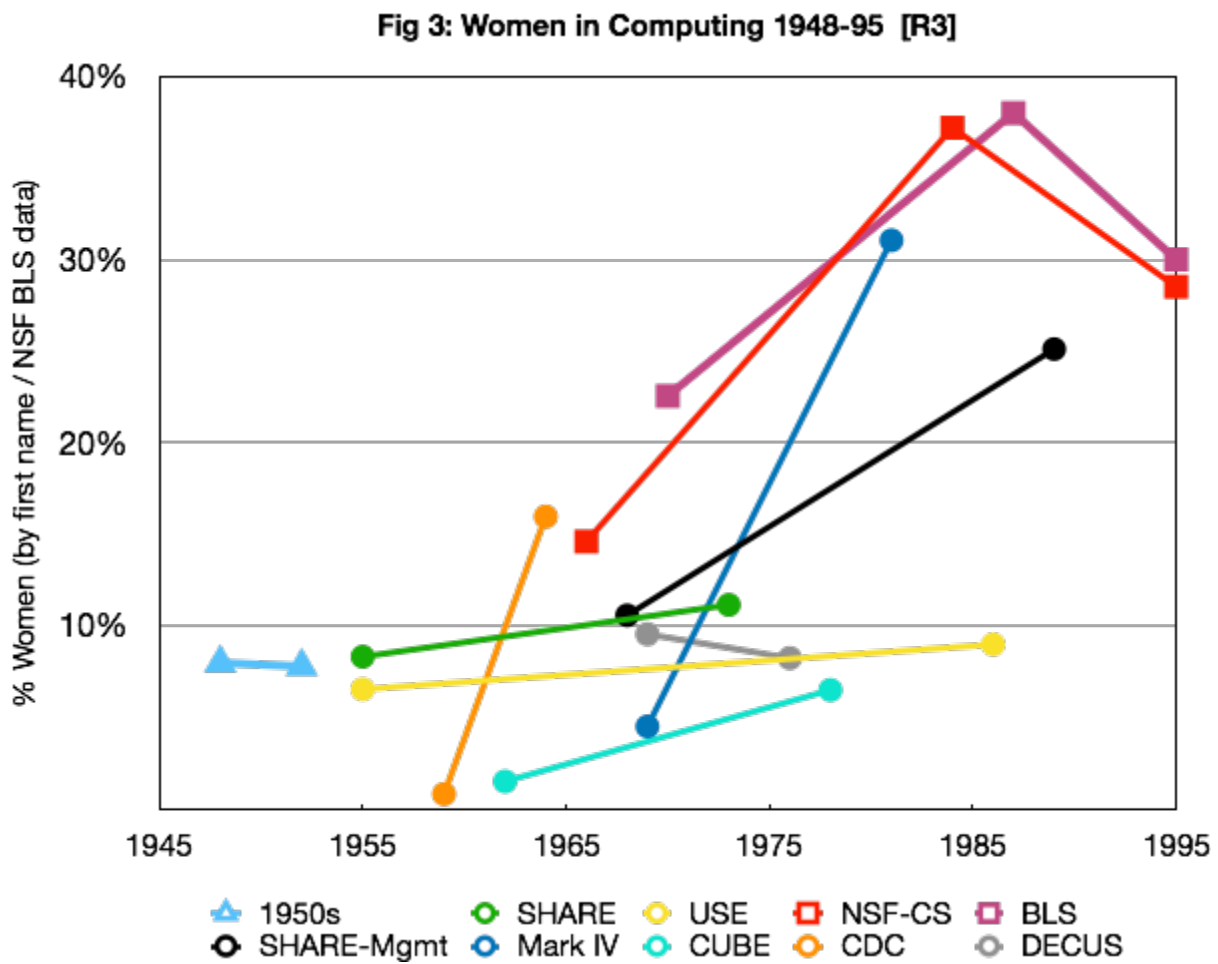

[v] *Business Automation* in 1960 found women were less than 15% of programmers; in its 1971 survey (N=600,000), women were "14% of systems analysts and 21% of computer programmers."[27]

[vi] See photos of attendees at NMAA (1951), ACM (various), and *company-wide* photographs from Control Data (1962, 1966, 1982) at web.archive.org/web/20201003135659/http://www.cbi.umn.edu/images/index.html

At least three distinct periods may be discerned. *First:* From 1948 through around 1960, women were a numerically small proportion of the computing community (ranging from 0 to around 10%). There is no systematic data—here or elsewhere—that women were anything like 30 to 50 percent of the skilled white-collar computing workforce, until the 1980s. Growth was modest (see 'USE' [slope = 0.0008]). The apparent sharp growth in CDC [N=371] reflects two years 1959-60 with *zero* women; data for Figure 1's '1950s' is not a proper time-series. *Second:* From the 1960s through the 1970s, women in computer-user groups grew steadily if slowly to reach roughly 12 to 20% (see 'CUBE' and 'SHARE' [slopes = 0.0031, 0.0016]). Women were entering computing during these years—despite the linear model's speculation about them leaving. (The only time-series showing any downward drift is DECUS in 1968, 1972, 1976 [N=2,116] easing from 9.5 to 8.2% women.) Subsequent data through the mid-1980s suggest accelerating growth in women's participation in computing (see 'SHARE-Mgmt' and 'Mark IV' [slopes = 0.007 and 0.022]). Women attending USE grew to reach 15% in the mid-1980s; women officer-managers of SHARE grew to 26% in 1989; and women attending Mark IV conferences grew to 30% in 1980. These data are consistent with the U.S. Census reporting 22.5% women in the computing workforce (1970) and with the peak years for women's participation in the mid-1980s.[22] *Third:* women indeed left computing—after the peak in the mid-1980s—and this is what has persisted to the present. According to the CRA Taulbee survey, women's share of computer-science bachelor's degrees fell to 11.2% (2009). The U.S. Census American Community Survey reported women constituted 27% of the computing workforce (in 2011), a precipitous drop from the mid-1980s peak of 38%.[38]

These three periods demonstrate a non-linear dynamic for gender bias in computing. Instead of one question based on conjecture—"when" did women leave computing?—we now face distinct data-driven research questions. How did women establish a significant presence in the nascent high-skilled computer field in the 1950s? Men solidly dominated the fields that early computing drew on most heavily, such as engineering,[vii] physics, and mathematics;[viii] and yet computing women took up positions of responsibility and leadership such as Frances Holberton (née Snyder), Grace Hopper, Mina Rees, and many others. Why was women's growth in the computing workforce steady although slow through the mid-1960s? What attracted so many women into computing just as it professionalized during roughly 1965-85? Computing among scientific and technical fields stood out for its expanding hospitality to women during these two decades, and we should be alert for useful lessons. And, finally, how to understand the exodus of women beginning in the late-1980s that afflicts computing through today?

---

[vii] Bix writes, "As late as the 1960s, women still made up less than 1 percent of students studying engineering in the United States."[7] Available data are thin or non-existent for women in specific engineering or science *workforces*; many studies make estimates from *educational* data.

[viii] Mathematics prior to 1940 was distinctly open to women, who gained 14% of the field's Ph.D's.[23] But during 1945–1960 the number of men gaining math PhD's roughly tripled; while women experienced stasis in numbers and decline in participation (falling to 4.6–9.3% of total math PhD's).[30,44]

**Research for the future**

Further research is necessary to address these new questions, but it's clear the worrisome sea-change in computing during the late 1980s and 1990s accompanied dramatic cultural shifts. These include the rise of personal computing, gendered avatars in computer gaming, and the media's lionization of male "nerds." The nerd image, which had been previously ambiguous, flexible, and rhetorically situated distant from power, "gets rehabilitated and partially incorporated into hegemonic masculinity" beginning in the 1980s.[34] (Hegemonic masculinity can be defined as the "configuration of gender practice [that] guarantees [or is taken to guarantee] the dominant position of men and the subordination of women."[17]) Popular media such as "Revenge of the Nerds" (1984) and "Triumph of the Nerds" (1996) sharpened the nerd image as a computing male. And nerds became allied with power. *Wired* magazine offered up Nicholas Negroponte, Stewart Brand, George Gilder, and John Perry Barlow in the 1990s. "Wired is about the most powerful people on the planet today—the Digital Generation," stated its cofounder. Bill Gates graced its cover five times in 15 years (and later gained a sixth with Mark Zuckerberg).[37,55,56] Today, many researchers target computing's gender-slanted culture, ingrained stereotypes, and associated public images as promising sites for positive intervention.[15,16,21,31,33]

The labor-intensive research method reported here might be automated by linking meeting and membership records with the SSA dataset.[32] As a pilot, I analyzed the 1949 ACM roster (N=435) in two ways. First, I did manual spreadsheet tallies of listed individuals as woman's, man's, initials-only, or gender-ambiguous name; as usual, I resolved gender-unclear names though contextual-archival linking or the SSA dataset. Second, I drew on the SSA dataset (year-of-birth = 1925) to directly compute the gender probabilities of each name. All but three "male" names (n=160) had 95% or greater probability of being male. Noel (91%), Francis (90%), and Jan (45%) were the exceptions; in this instance, it was Jan Rajchman, the noted RCA Laboratory engineer and IAS computer designer. Only one "female" name (n=27) had less than a 99% probability of being female. Jean is a US woman's name (97.5%) but a Francophone male name; the 'Jeans' from (e.g.) Hydro-Québec attending these meetings indicate the need for contextual knowledge to correctly infer gender. In addition, 15 first names did not appear in the SSA dataset and were set aside. A weighed sum of the "male" and "female" name probabilities directly computed with one minor adjustment (resolving 8 "initials-only" ACM members who were well-known men, namely JH Boekhoff, JG Brainerd, H Campaign, JJ Eachus, RW Hamming, CC Hurd, CC Gotlieb, and MV Wilkes) predicted that 8.52% of that year's ACM members were women, close to the manually tabulated 8.55% women. APIs exist for inferring gender from first names,[42,49] and some may deal with temporal changes in ascribed gender for such names as "Robin" or "Leslie" or even international names beyond the US-based SSA dataset.[48]

Women's advances in the computing profession from the 1960s through the 1980s deserve special scrutiny today; in these years, computing was attractive to literally thousands of women programmers, systems analysts, database specialists, and middle managers. It is a mistaken notion that computing was somehow "made masculine" during these years when, in fact, women were flooding into the profession—attending professional meetings, participating in computer-user groups, and earning an increasing share of computer-science bachelor's degrees. The "making programming masculine" thesis has unwittingly obscured the very years when women found computing to be an exciting field where their technical talents could be actively exercised and professionally rewarded.[1,10,28,40,57] Recent retirements of top women executives at IBM, HP, and Xerox underscore the peak years of the 1980s when these women launched computing careers and when the field was nearly 40% women.[ix]

More detailed gender-analysis of membership lists and conference attendees of ACM's numerous SIGs could shed light on which branches of computer science evinced greater or lesser openness to women's participation. Some branches of computer security had especially noteworthy women's leadership. For example, pioneering intrusion-detection research was led by Dorothy Denning, Teresa Lunt, Debra Anderson, Rebecca Bace, and others.[40,58] HCI has focused research attention on gender.[6,11,54] Recent findings suggest gender bias may be endemic in the *content* of machine learning, as expressed in the meme "Man is to Computer Programmer as Woman is to Homemaker."[4,8,35] Data beyond user groups is desirable. ACM members likely possess SIG records that could advance our understanding of the dynamics of gender bias in computing. ACM's History Committee recently launched a SIG-focused archiving initiative.[39] A large-scale data-gathering effort could empirically analyze what computing did right during the 1960s-1980s—focusing on specific SIGs and subfields—as well as what went wrong during the 1990s and beyond. If the preliminary research reported here is extended, perhaps the hard problem of gender bias in computing can be made tractable.

---

[ix] See geekfeminism.wikia.org/wiki/List_of_women_executives_at_tech_companies and www.fastcompany.com/1139328/women-tech-executives

**Thomas J Misa** is Past President of the Society for the History of Technology (2021-22) and editorial board member for ACM Books (2013-present). He directed the Charles Babbage Institute (2006-17) at the University of Minnesota and chaired of the ACM History Committee (2014-16). Research supported by Alfred P. Sloan Foundation grant G-B2014-07.